\documentclass[a4paper,twoside]{article}

\usepackage{epsfig}
\usepackage{subfigure}
\usepackage{calc}
\usepackage{amssymb}
\usepackage{amstext}
\usepackage{amsmath}
\usepackage{amsthm}
\usepackage{multicol}
\usepackage{pslatex}
\usepackage{apalike}
\usepackage{SCITEPRESS}
\usepackage[small]{caption}

\usepackage{balance}

\usepackage{ifxetex,ifluatex}
\ifxetex
  \usepackage[setpagesize=false, 
              unicode=false, 
              xetex]{hyperref}
\else
  \usepackage[unicode=true]{hyperref}
\fi
\hypersetup{breaklinks=true,
            bookmarks=true,
            pdfauthor={},
            pdftitle={Heart Rate Monitoring as an Easy Way to Increase Engagement in Human-Agent Interaction},
            colorlinks=true,
            citecolor=blue,
            urlcolor=blue,
            linkcolor=magenta,
            pdfborder={0 0 0}}
\usepackage[T1]{fontenc}
\usepackage{ifxetex,ifluatex}
\ifnum 0\ifxetex 1\fi\ifluatex 1\fi=0 
  \usepackage[utf8]{inputenc}
\fi

\usepackage{comment}

\makeatletter
\def\maxwidth{\ifdim\Gin@nat@width>\linewidth\linewidth
\else\Gin@nat@width\fi}
\makeatother
\let\Oldincludegraphics\includegraphics
\renewcommand{\includegraphics}[1]{\Oldincludegraphics[width=\maxwidth]{#1}}

\def \citep {\cite}

\begin{document}

\excludecomment{quote}

\title{Heart Rate Monitoring as an Easy Way to Increase Engagement in
Human-Agent Interaction}


\author{
\authorname{
      Jérémy Frey
                  \unskip\sup{1,2,3}
  \unskip
  }
\affiliation{\sup{1}Univ. Bordeaux, LaBRI, UMR 5800, F-33400 Talence, France}
\affiliation{\sup{2}CNRS, LaBRI, UMR 5800, F-33400 Talence, France}
\affiliation{\sup{3}INRIA, F-33400 Talence, France}
\email{
jeremy.frey@inria.fr}
}


\abstract{Physiological sensors are gaining the attention of manufacturers and
users. As denoted by devices such as smartwatches or the newly released
Kinect 2 -- which can covertly measure heartbeats -- or by the
popularity of smartphone apps that track heart rate during fitness
activities. Soon, physiological monitoring could become widely
accessible and transparent to users. We demonstrate how one could take
advantage of this situation to increase users' engagement and enhance
user experience in human-agent interaction. We created an experimental
protocol involving embodied agents -- ``virtual avatars''. Those agents
were displayed alongside a beating heart. We compared a condition in
which this feedback was simply duplicating the heart rates of users to
another condition in which it was set to an average heart rate. Results
suggest a superior social presence of agents when they display feedback
similar to users' internal state. This physiological
``similarity-attraction'' effect may lead, with little effort, to a
better acceptance of agents and robots by the general public.}

\onecolumn

\maketitle

\normalsize

\vfill

\section{INTRODUCTION}\label{introduction}

\noindent Covert sensing of users' physiological state is likely to open
new communication channels between human and computers. When
anthropomorphic characteristics are involved -- as with embodied agents
-- mirroring such physiological cues could guide users' preferences in a
cheap yet effective manner.

One aspect of human-computer interaction (HCI), albeit difficult to
account for, lies in users' engagement. Engagement may be seen as a way
to increase performance, as in the definition given by
\citep{Matthews2002} for task engagement: an ``effortful striving
towards task goals''. In a broader acceptation, the notion of engagement
is also related to fun and accounts for the overall user experience
\citep{Mandryk2006}. Several HCI components can be tuned to improve
engagement. For example, content and challenge need to be adapted and
renewed to avoid boredom and maintain users in a state of flow
\citep{Berta2013}. It is also possible to study interfaces:
\citep{Karlesky2014} use tangible interactions in surrounding space to
spur engagement and creativity. When the interaction encompasses
embodied agents -- either physically (i.e., robots) or not (on-screen
avatars) -- then anthropomorphic characteristics can be involved to seek
better human-agent connections.

\begin{quote}
also for boredom: Nacke2010
\end{quote}

Following the affective computing outbreak \citep{Picard1995}, studies
using agents that possess human features in order to respond to users
with the appropriate emotions and behaviors began to emerge.
\citep{Prendinger2004} created an ``empathic'' agent that serves as a
companion during a job interview. While playing on empathy to engage
users more deeply into the simulation was conclusive, the difficulty
lies in the accurate recognition of emotions. Even using physiological
sensors, as did the authors with galvanic skin response and
electromyography, no signal processing could yet reach an accuracy of
100\%, even on a reduced set of emotions -- see \citep{Lisetti2004} for
a review.

Humans are difficult to comprehend for computers and, still, humans are
more attracted to others -- human or \emph{machine} -- that match their
personalities \citep{Lee2003}. This finding is called
``similarity-attraction'' in \citep{Lee2003} and was tested by the
authors by matching the parameters of a synthesized speech (e.g.,
paralinguistic cues) to users, whenever they were introverted or
extroverted. An analogous effect on social presence and engagement in
HCI has been described as well in \citep{Reidsma2010a}, this time under
the name of ``synchrony'' and focusing on nonverbal cues (e.g.,
gestures, choice of vocabulary, timing, \ldots{}). Unfortunately, being
somewhat linked to a theory of mind, such improvements lean against
tedious measures, for instance psychological tests or recordings of
users' behaviors. What if the similarity-attraction could be effective
with cues that are much simpler and easier to set up?

Indeed, at a lower level of information, \citep{Slovak2012} studied how
the display of heart rate (HR) could impact social presence during
human-human interaction. They showed that, without any further
processing than the computation of an average heartbeat, users did
report in various contexts being closer or more connected to the person
with whom they shared their HR. We wondered if a similar effect could be
obtained between a human and a machine. Moreover, we anticipated the
rise of devices that could covertly measure physiological signals, such
as the Kinect 2, which can use its cameras (color and infrared) to
compute users' HRs -- the use of video feeds to perform volumetric
measurements of organs is dubbed as ``photoplethysmography''
\citep{Kranjec2014}.

Consequently, we extended on the theory and \textbf{we hypothesized that
users would feel more connected toward an embodied agent if it displays
a heart rate similar to theirs, \emph{even if users do not realize that
their own heart rates are being monitored}.}

By relying on a simple mirroring of users' physiology, we elude the need
to test users' personality \citep{Lee2003} or to process -- and
eventually fail to recognize -- their internal state
\citep{Prendinger2004}. Creating agents too much alike humans may
provoke rejection and deter engagement due to the uncanny valley effect
\citep{MacDorman2005}. Since we do not emphasize the link between users'
physiological cues and the feedback given by agents, we hope to prevent
such negative effect. The similary-attraction applied to physiological
data should work at an almost subconscious level. Furthermore, implicit
feedback makes it easier to improve an existing HCI. As a matter of
fact, only the feedback associated with the agent has to be added to the
application; feedback that can then take a less anthropocentric form --
e.g., see \citep{Harrison2012} for the multiple meanings a blinking
light can convey and \citep{Huppi2003} for a use case with
breathing-like features. Ultimately, our hypothesis proved robust, it
could benefit to virtually any human-agent interaction, augmenting
agent's social presence, engaging users.

The following sections describe an experimental setup involving embodied
agents that compares two within-subject conditions: one condition during
which agents display heartbeats replicating the HR of the users, and a
second condition during which the displayed heartbeats are not linked to
users. Our main contribution is to show first evidence that displaying
identical heart rates makes users more engaged toward agents.

\section{EXPERIMENT}\label{experiment}

\label{sec:xp_description}

\noindent The main task of our HCI consisted in listening to embodied
agents while they were speaking aloud sentences extracted from a text
corpus, as inspired by \citep{Lee2003}. When an agent was on-screen, a
beating heart was displayed below it and an audio recording of a heart
pulse was played along each (fake) beat. This feedback constituted our
first within-subject factor: either the displayed HR was identical to
the one of the subject (``human'' condition), either it was set at an
average HR (``medium'' condition). The HR in the ``medium'' condition
was ranging from 66 to 74 BPM (beats per minute), which is the grand
average for our studied population \citep{Agelink2001}.

Agents possessed some random parameters: their gender (male or female),
their appearance (6 faces of different ethnic groups for each gender),
their voice (2 voices for each gender) and the voice pitch. Those
various parameters aimed at concealing the true independent variable.
Had we chosen a unique appearance for all the agents, subjects could
have sought what was differentiating them. By individualizing agents we
prevented subjects to discover that ultimately we manipulated the HR
feedback. To make agents look more alive, their eyes were sporadically
blinking and their mouths were animated while the text-to-speech system
was playing.

\begin{quote}
with strong emotions, more likely to get attach? true with human (ref
needed), but then it's another study.
\end{quote}

In order to elicit bodily reactions, we chose sentences for which a
particular valence has been associated with, and, as such, that could
span a wide range of emotions. Valence relates to the hedonic tone and
varies from negative (e.g., sad) to positive (e.g., happy) emotions
\citep{Picard1995}. HR has a tendency to increase when one is
experiencing extreme pleasantness, and to decrease when experiencing
unpleasantness \citep{Winton1984}.

\begin{figure}[htbp]
\centering
\includegraphics{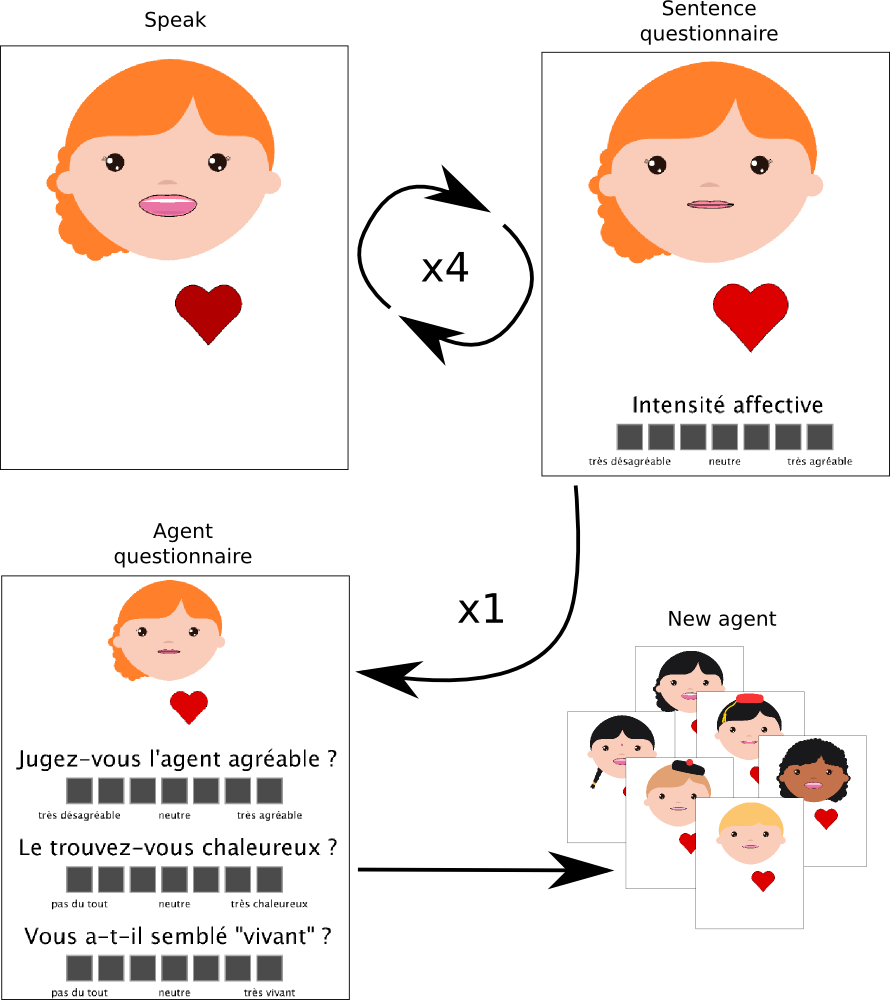}
\caption{Procedure during the ``disruptive'' session: subjects rate the
valence of each one of the sentences spoken by an agent. After 4
sentences, they rate agent's social presence (3 items). Then a new agent
appears. 20 agents, average time per agent $\approx$
62.2s.\label{fig:xp1}}
\end{figure}

Our experiment was split in two parts (second within-subject factor).
During the first session, called ``disruptive'' session (see Figure
\ref{fig:xp1}), subjects had to rate each sentence they heard on a
7-point Likert scale according to valence they perceived (very
unpleasant to very pleasant). Sentences came from newspapers. A valence
(negative, neutral or positive) was randomly chosen every 2 sentences.
Every 4 sentences, subjects had to rate the social presence of the
agent. Then a new randomly generated agent appeared, for a total of 20
agents, 10 for each ``human''/``medium'' condition.

\begin{figure}[htbp]
\centering
\includegraphics{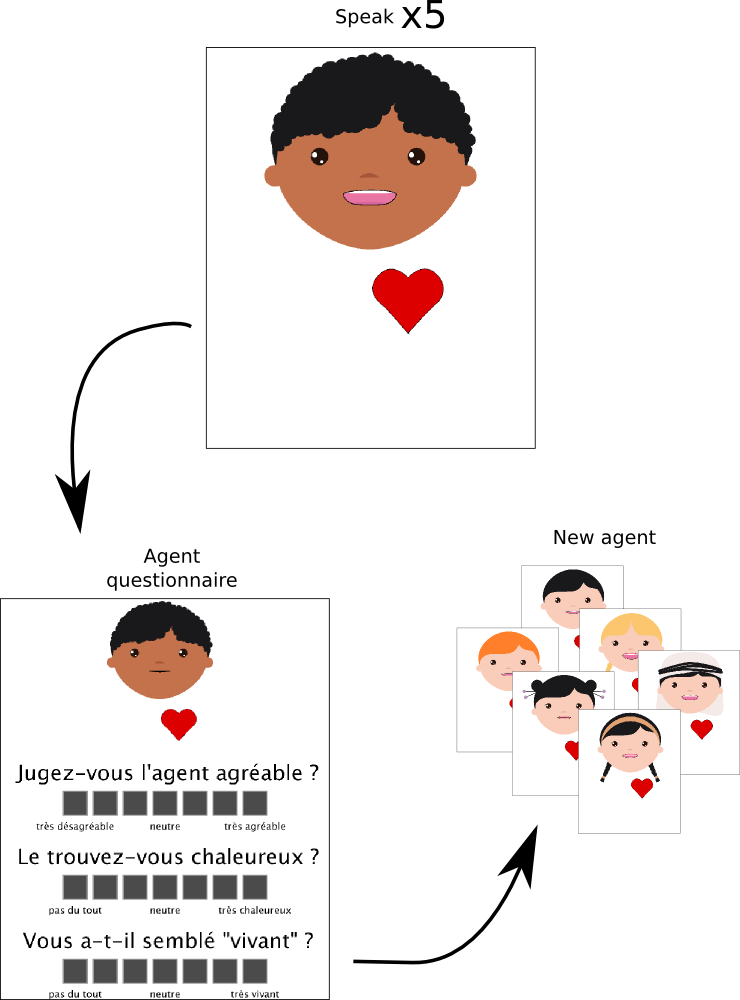}
\caption{Procedure during the ``involving'' session: subjects rate
agent's social presence after it recited all its sentences. Then a new
agent appears, continuing the tale. 23 agents, average time per agent
$\approx$ 46.6s.\label{fig:xp2}}
\end{figure}

As opposed to the first part, during the second part of the experiment,
called ``involving'' session, sentences order was sequential (see Figure
\ref{fig:xp2}). Agents were in turns narrating a fairy tale. Subjects
did not have to rate each sentence's valence, instead they only rated
the social presence of the agents. To match the length of the story,
agents were shuffled every 6 sentences and there were 23 agents in
total, 12 for the ``human'' condition, 11 for the ``medium'' condition.

Because of its distracting task and the nature of its sentences, the
first part was more likely to disrupt human-agent connection; while the
second part was more likely to involve subjects. This let us test the
influence of the relation between users and agents on the perception of
HR feedback. We chose not to randomize sessions order because we
estimated that putting the ``disruptive'' session last would have made
the overall experiment too fatiguing for subjects. A higher level of
vigilance was necessary to sustain its distracting task and series of
unrelated sentences. Subjects' cognitive resources were probably higher
at the beginning of the experiment.

We created a 2 (HR feedback: ``human'' \emph{vs} ``medium'' condition) x
2 (nature of the task: ``disruptive'' \emph{vs} ``involving'' session)
within-subject experimental plan. Hence our two hypothesis. \textbf{H1}:
Hear rate feedback replicating users' physiology increases the social
presence of agents. \textbf{H2}: This effect is more pronounced during
an interaction involving more deeply agents.

\subsection{Technical description}\label{technical-description}

Most of the elements we describe in this section, hardware or software,
come from open source movements, for which we are grateful. Authors
would also like to thank the artist who made freely available the
graphics on which agents are based\footnote{\url{http://harridan.deviantart.com/}}.
All code and materials related to the study are freely available at
\url{https://github.com/jfrey-phd/2015_phycs_HR_code/}.

\begin{quote}
dead website for artist? http://waderadesign.com/ Acknowledge in
dedicated section?
\end{quote}

\subsubsection{Hardware}\label{hardware}

\begin{figure}[htbp]
\centering
\includegraphics{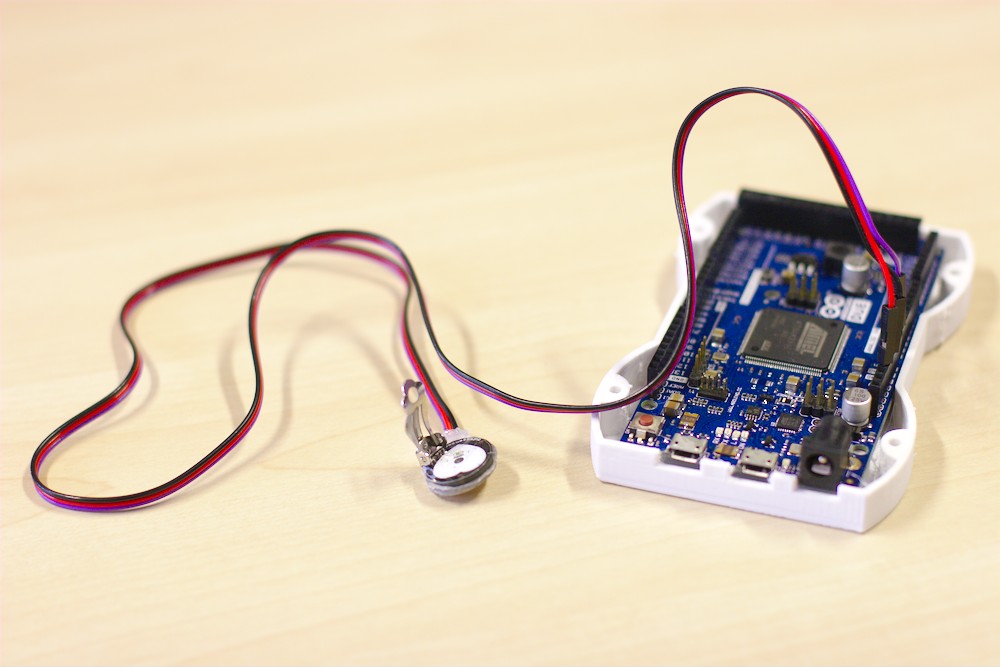}
\caption{BVP (blood volume pulse) sensor measuring heartbeats, connected
to an Arduino Due.\label{fig:sensor}}
\end{figure}

We chose to use a BVP (blood volume pulse) sensor to measure HR,
employing the open hardware Pulse Sensor\footnote{\url{http://pulsesensor.myshopify.com}}
(see Figure \ref{fig:sensor} for a closeup). It assesses blood flow
variations by emitting a light onto the skin and measuring back how
fluctuates the intensity of the reflected light thanks to an ambient
light photo sensor. Each heartbeat produces a characteristic signal.
This technology is cheap and easy to implement. While it is less
accurate than electrocardiography (ECG) recordings, we found the HR
measures to be reliable enough for our purpose. Compared to ECG, BVP
sensors are less intrusive and quicker to install -- i,e,. one sensor
around a finger or on an earlobe instead of 2 or 3 electrodes on the
chest. In addition, as far as general knowledge is concerned, BVP
sensors are less likely to point out the exact nature of their measures.
This ``fuzziness'' is important for our experimental protocol, as we
want to be as close as possible to the real-life scenarios we foresee
with devices such as the Kinect 2, where HR recordings will be
transparent to users.

The BVP sensor was connected to an Arduino Due\footnote{\url{http://arduino.cc/}}
(see Figure \ref{fig:sensor}). Arduino boards have become a
well-established platform for electrical engineering. The Due model
comes forward due to its 12 bits resolution for operating analog
sensors. The program uploaded into the Arduino Due was feeding the
serial port with BVP values every 2ms, thus achieving a 500Hz sampling
rate.

\begin{quote}
give exact laptop references? there's two for experimenter
laptop\ldots{} alienware m14x for subject. the other, mainly lenovo
t61p, for 2 subjects alienware ?? (the big guy)
\end{quote}

\begin{quote}
skipped headset. good enough (especially deep sounds) and\ldots{} comes
from merchandising, footloose movie :D
\end{quote}

Two computers were used. One, a 14 inches screen laptop, was dedicated
to the subject and ran the human-agent interaction. This computer was
also plugged to the Arduino board to accommodate sensor's cable length.
A second laptop was used by the experimenter to monitor the experiment
and to detect heartbeats. Computers were connected through an ethernet
cable (network latency was inferior to 1ms).

\subsubsection{Software and signal
processing}\label{software-and-signal-processing}

\begin{quote}
do not talk about stimulations, useless in here (but a big piece of work
to trigger them all and sync everything!)
\end{quote}

Computers were running Kubuntu 13.10 operating system. The software on
the client side was programmed with Processing framework\footnote{\url{http://www.processing.org/}},
version 2.2.1. Data acquired from the BVP sensor was streamed to the
local network with ser2sock\footnote{\url{https://github.com/nutechsoftware/ser2sock}}.
This serial port-to-TCP bridge software allowed us to reliably process
and record data on our second computer. OpenViBE \citep{Renard2010}
version 0.18 was running on the experimenter's computer to process BVP.

\begin{quote}
short version for BPM, there's timeout control, cutoff interval, +/-
10\% max variation to avoid glitches
\end{quote}

\begin{figure}[htbp]
\centering
\includegraphics{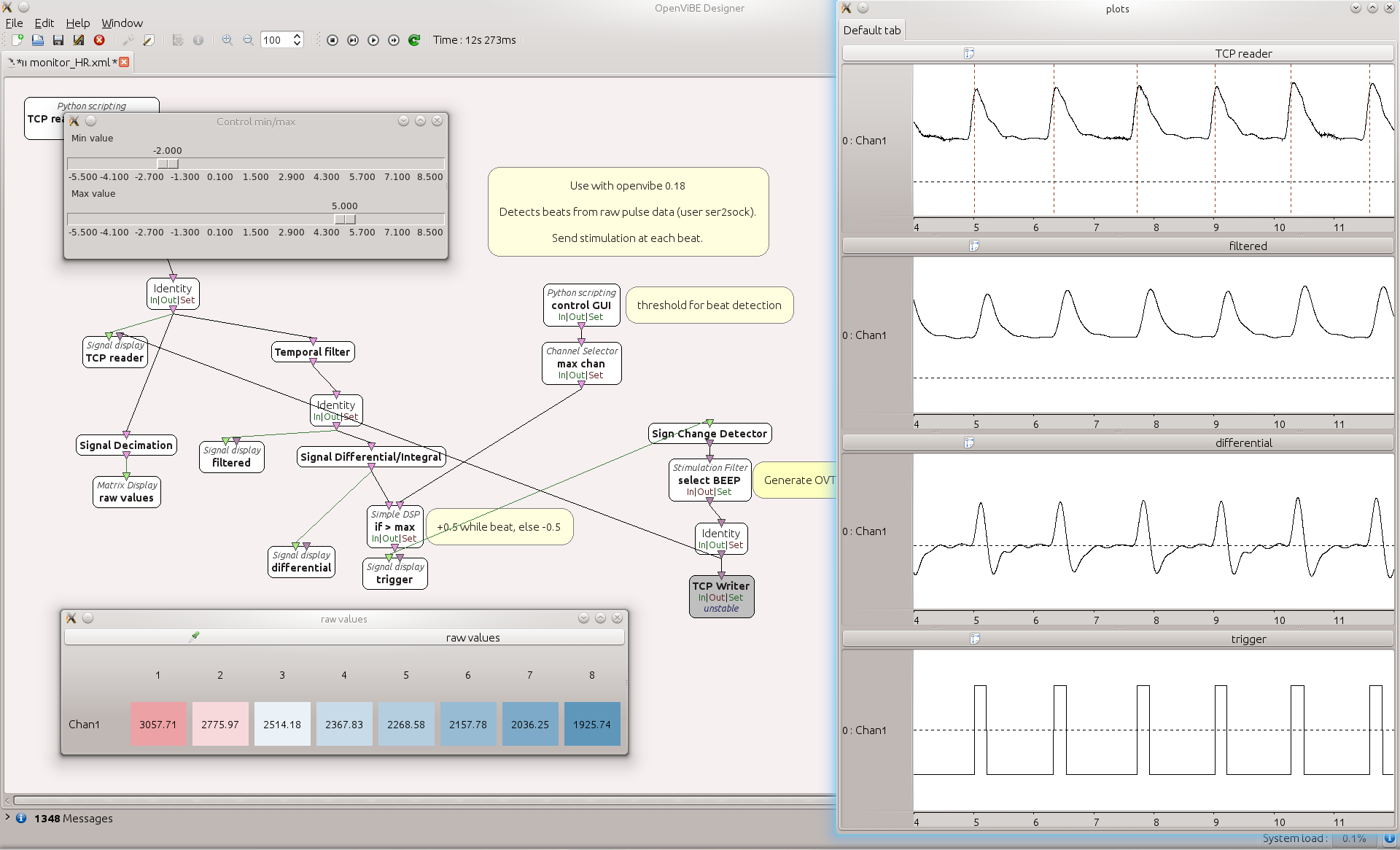}
\caption{Signal processing of the BVP sensor with OpenViBE. A low-pass
filtered and a first-derivative are used to detect
heartbeats.\label{fig:ov}}
\end{figure}

Within OpenViBE the BVP values were interpolated from 500 to 512Hz to
ease computations. The script which received values from TCP was
downsampling or oversampling packets' content to ensure synchronization
and decrease the risk of distorted signals due to network or computing
latency. A 3Hz low-pass filter was applied to the acquired data in order
to eliminate artifacts. Then a derivative was computed. Since a
heartbeat provokes a sudden variation of blood flow, a pulsation was
detected when the signal exceeded a certain threshold. This threshold
was set during installation: values too low could produce false
positives due to remaining noise, and values too high could skip
heartbeats. Eventually a message was sent. See figure \ref{fig:ov} for
an overview of the signal processing.

\begin{quote}
Using raw TCP communication and Ethernet cable, delays, for signal or
stimulation where negligible (inferior to 1ms). Even the bottleneck
induced by a 60 FPS of the main program couldn't impact our processing.
\end{quote}

Once the main program received a pulse message, it computed the HR from
the delay between two beats. This value was passed over the engine
handling the HR feedback during the ``human'' condition. We purposely
created an indirection here -- using BPM values in separate handlers
instead of triggering a feedback pulse as soon as a heartbeat was
detected -- in order to suit our experimental protocol to devices that
could only average HR over a longer time window (e.g., fitness HR
monitor belts). It should be easier to replicate our results without the
need to synchronize precisely feedback pulses with actual heartbeats.

\begin{quote}
espeak 1.47.11, MBROLA 3.01h
\end{quote}

The TTS (text-to-speech) system comprised two applications.
eSpeak\footnote{\url{http://espeak.sourceforge.net/}} was used to
transform textual sentences into phonemes and MBROLA\footnote{\url{http://tcts.fpms.ac.be/synthesis/mbrola.html}}
to synthesize phonemes and produce an actual voice. The TTS speed was
controlled by eSpeak (120 word per minutes), as well as the pitch
(between 65 and 85, values higher than the baseline of 50 to match the
teenage appearance of the agents). The four voices (2 male and 2 female,
``fr1'' to ``fr4'') were provided by the MBROLA project. Sentences'
valence did not influence speech synthesis.

\subsection{Text corpuses}\label{text-corpuses}

During the first part of the experiment (i.e., the ``disruptive''
session) sentences were gathered from archives of a french-speaking
newspaper. These data were collated by \citep{Bestgen2004}. Sentences
were anonymized, e.g., names of personalities were replaced by generic
first names. A panel of 10 judges evaluated their emotional valence on a
7-point Likert scale. The final scores were produced by averaging those
10 ratings. We split the sentences in three categories: unpleasant
(scores between $[-3;-1[$, e.g., a suspect was arrested for murder),
neutral (between $[-1;1]$) and pleasant (between $]1;3]$, e.g., the
national sport team won a match) -- see section
\ref{sec:xp_description}.

The sentences of the second part (i.e., the ``involving'' session) come
from the TestAccord Emotion database \citep{LeTallec2011}. This database
originates from a fairy tale for children -- see \citep{Wright2008} for
an example of storytelling as an incentive for empathy. We did not
utilize \emph{per se} the associated valences (average of a 5-point
Likert scale across 27 judges for each sentence), but as an indicator it
did help us to ensure the wide variety of the carried emotions. For
instance, deaths or bonding moments are described during the course of
the tale.

It is worth noting that when the valence of these corpuses has been
established, sentences were presented in their textual form, not through
a TTS system.

\subsection{Procedure}\label{procedure}

\begin{quote}
talk about the experiment sometimes going at home?
\end{quote}

The overall experiment took approximately 50 minutes per subject. 10
French speaking subjects participated in the experiment; 5 males, 5
females, mean age 30.3 (SD=8.2). The whole procedure comprised the
following steps:

\begin{enumerate}
\def\labelenumi{\arabic{enumi}.}
\item
  Subjects were given an informed consent and a demographic
  questionnaire. While they filled the forms, the equipment was set up.
  Then we explained to them the procedure of the experiment. We
  emphasized the importance of the distraction task (i.e., to rate
  sentences' valence) and explained to the subjects that we were
  monitoring their physiological state, without further detail about the
  exact measures. $\approx$ 5 min.
\item
  The BVP sensor was placed on the earlobe opposite to the dominant
  hand, so as not to impede mouse movements. Right after, the headset
  was positioned. We ensured that subjects felt comfortable, in
  particular we checked that the headset wasn't putting pressure on the
  sensor. We started to acquire BVP data and adjusted the heartbeat
  detection. $\approx$ 2 min.
\item
  A training session took place. We started our program with an
  alternate scenario, adjusting the audio volume to subjects' taste.
  Both parts of the experiment occurred, but with only two agents and
  with a dedicated set of sentences. This way subjects were familiarized
  with the task and with the agents -- i.e., with their general
  appearance and with the TTS system. During this overview, so as not to
  bias the experiment, ``human'' and ``medium'' conditions were replaced
  with a ``slow'' HR feedback (30 BPM) and a ``fast'' HR feedback (120
  BPM). Once subjects reported that they understood the procedure and
  were ready, we proceeded to the experiment. $\approx$ 5 min.
\end{enumerate}

\begin{figure}[htbp]
\centering
\includegraphics{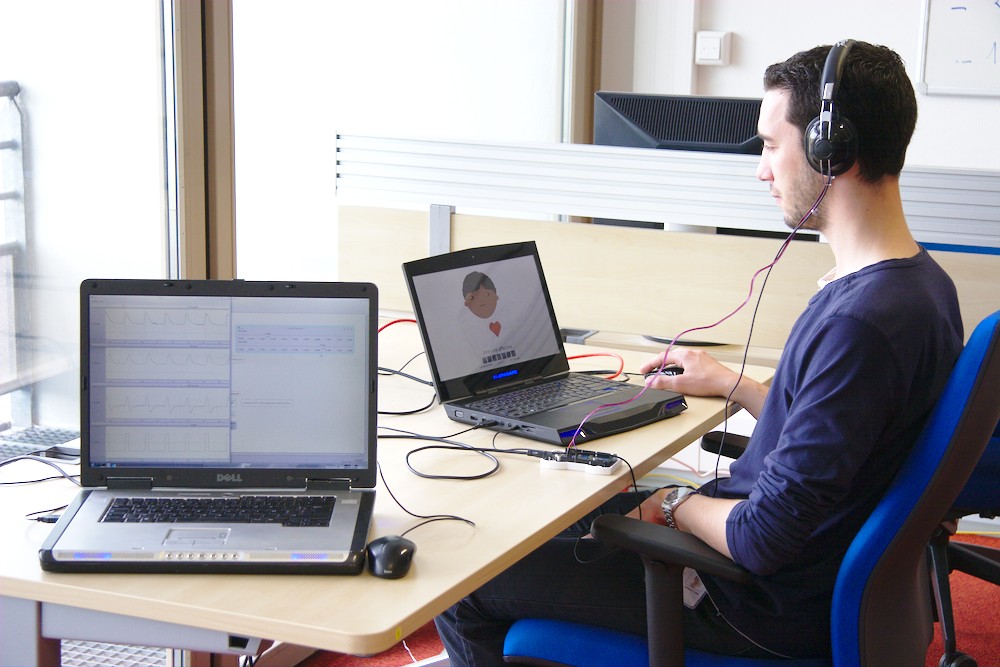}
\caption{Our experimental setup. A BVP sensor connects subject's earlobe
to the first laptop, where the human-agent interaction takes place.
Subject is wearing a headset to listen to the speech synthesis. A second
laptop is used by the experimenter to monitor heartbeats
detection.\label{fig:setup}}
\end{figure}

\begin{enumerate}
\def\labelenumi{\arabic{enumi}.}
\setcounter{enumi}{3}
\itemsep1pt\parskip0pt\parsep0pt
\item
  We ran the experiment, as previously described. First the
  ``disruptive'' session (80 sentences, 20 agents, $\approx$ 22 min),
  then the ``involving'' session (138 sentences, 23 agents, $\approx$ 17
  min). We were monitoring the data acquired from the BPV sensor and
  silently adjusted the hearbeat detection through OpenViBE if needed --
  rarely, a big head movement could slightly move the sensor and modify
  signal amplitude. Figure \ref{fig:setup} illustrates our setup.
  $\approx$ 40 min.
\end{enumerate}

The newspapers sentences being longer than the ones forming the fairy
tale, agents on-screen time varied between both parts. Agents mean
display time during the first part was 62.2s, during the second part it
was 46.6s.

\subsection{Measures}\label{measures}

We computed a score of social presence for each agent, averaged from the
7-point Likert scales questionnaires presented to the subjects before a
new agent were generated. This methodology was validated with spoken
dialogue systems by \citep{Moller2007}. This score was composed of 3
items, consistent with ITU guidelines \citep{ITU2003}. Translated to
English, the items were: ``Do you consider that the agent is pleasant?''
(``very unpleasant'' to ``very pleasant''); ``Do you think it is
friendly?'' (``not at all'' to ``very friendly''); ``Did it seem
`alive'?'' (``not at all'' to ``much alive'').

\subsection{Results}\label{results}

\begin{quote}
if ``greater'' hypothesis used for wilcoxon test: p \textless{} 0.01
overall and, with p \textless{} 0.05, ``When we looked deeper at the
data, these results came out the same for the items of the social
presence score taken separately.''
\end{quote}

We compared agents' social presence scores between the ``human'' and the
``medium'' conditions for each part. Statistical analyses were performed
with R 3.0.1. The different scores were comprised between 0 (negative)
and 6 (positive), 3 corresponding to neutral.

A Wilcoxon Signed-rank test showed a significant difference (p
\textless{} 0.05) during the ``disruptive'' session (means 3.29
\emph{vs} 2.91) but no significant difference (p = 0.77) during the
``involving'' session (means: 3.30 \emph{vs} 3.34). H1 is verified while
H2 cannot be verified. Besides, when we analyzed further the data, we
found no significant effect (p = 0.27) of the ``human''/``medium''
factor on the valence scores attributed to the sentences during the
``disruptive'' session (means: 3.06 \emph{vs} 2.91).

Subjects' HRs were a little higher than expected during the experiment:
mean $\approx$ 74.73 BPM (SD = 5.59); to be compared with the average 70
BPM set in the ``medium'' condition. We used Spearman's rank correlation
test to check whenever this factor could have influenced the results
obtained in the ``disruptive'' session. To do so, we compared subjects'
average HRs with the differences in social presence scores between
``human'' and ``medium'' conditions. There was not significant
correlation (p = 0.25).

\begin{quote}
TODO: compare social presence between first and second part?
\end{quote}

\begin{quote}
TODO: compute correlation between answers and ``ground truth''?
\end{quote}

\section{DISCUSSION}\label{discussion}

\noindent In the course of the ``disruptive'' session our main
hypothesis has been confirmed: users' engagement toward our HCI
increased when agents provided feedback mirroring their physiological
state. This result could not be explained by a preference for a certain
pace of the HR feedback. For instance, even though their HRs were higher
than average, subjects did not prefer agents of the ``human'' condition
because of faster heartbeats. Some of them did possess HRs lower than 70
BPM. The only other explanation lies in the difference of HR
synchronization between ``human'' and ``medium'' conditions.

Beside agents' social presence, similarity-attraction effect may
influence the general mood of subjects, as they had a slight tendency to
overrate sentences valence during ``human'' condition. It is interesting
to note that while the increase in social presence scores is not huge
(+13\%), it shifts the items from slightly unpleasant to slightly
pleasant.

Maybe the effect would have been greater in a less artificial situation.
Indeed, despite our experimental protocol, subjects reported afterwards
that the TTS system was sometimes hard to comprehend, which bothered
them on some occasions. It may have resulted in a task not involving
enough for the subjects to really ``feel'' the emotions carried by the
sentences.

Several reasons could explain why the effect appeared only during our
``disruptive'' session. During the first session agents were displayed
on a longer duration (+33\%) because of the longer sentences used in the
newspapers. The attraction toward a mirrored feedback could take time to
occur. In addition, because the task was less disruptive in the second
session, subjects were more likely to focus their attention on the
content (i.e., the narrative) instead of the interface (i.e., the
feedback). This could explain why they were less sensible to ambient
cues. Subject were less solicited during the ``involving'' session; we
observed that between agents questionnaires they often removed their
hands from the mouse, leaning back on the chair. Lastly, the
``involving'' session systematically occurred in second position. Maybe
the occurrence of the similarity-attraction effect is correlated to the
degree of users' vigilance.

As for subjects' awareness of the real goal of the study, during
informal discussions after the experiments, most of them confirmed that
they had no knowledge about the kind of physiological trait the sensor
was recording, and none of them realized that at some point they were
exposed to their own HR. This increases the resemblance of our
installation with a setup where HR sensing occurs covertly.

\section{CONCLUSION}\label{conclusion}

\noindent We demonstrated how displaying physiological signals close to
users could impact positively social presence of embodied agents. This
approach of ``ambient'' feedback is easier to set up and less prone to
errors than feedback as explicit as facial expressions. It does not
require prior knowledge about users nor complex computations. For
practical reasons we limited our study to a virtual agent. We believe
the similarity-attraction effect could be even more dramatic with
\emph{physically} embodied agents, namely robots. That said, other piece
of hardware or components of an HCI could benefit from such approach.
While its appearance is not anthropomorphic, the robotic lamp presented
by \citep{Gerlinghaus2012} behaves like a sentient being. Augmenting it
with physiological feedback, moreover when correlated to users, is
likely to increase its presence.

Further research is of course mandatory to confirm and analyze how the
similarity-attraction applies to human-agent interaction and to
physiological computing. The kind of feedback given to users need to be
studied. Are both audio and visual cues necessary? Does the look of the
measured physiological signal need to be obvious or could a heart pulse
take the form of a blinking light? In human-human interaction such
questions are more and more debated
\citep{Slovak2012};\citep{Walmink2014}. Obviously, one should check that
a physiological feedback does not \emph{diminish} user experience.
\citep{Lee2014} suggest it is not the case, but the comparison should be
made again with human-agent interaction.

Various parameters in human-agent interaction need to be examined to
shape the limits of the similarity-attraction effect: exposure time to
agents, nature of the task, involvement of users, and so on. Especially,
we suspect the relation between human and agent to be an important
factor. Gaming settings are good opportunities to try collaboration or
antagonism. Concerning users, some will perceive differently the
physiological feedback. As a matter of fact, interoception -- the
awareness of internal body states -- varies from person to person and
affects how we feel toward others \citep{Fukushima2011}. It will be
beneficial to record finely users reactions, maybe by using the very
same physiological sensors \citep{Becker2005}.

Finally, our findings should be replicated with other hardware. We used
lightweight equipment to monitor HR, yet devices such as the Kinect 2 --
if as reliable as BVP or ECG sensors -- will enable remote sensing in
the near future. But with the spread of devices that sense users'
physiological states, it is essential not to forgo ethics.

Measuring physiological signals such as HR enters the realm of privacy.
Notably, physiological sensors can make accessible to others data
unknown to self \citep{Fairclough2014}. Even though among a certain
population there is a trend toward the exposition of private data, if no
agreement is provided it is difficult to avoid a violation of intimacy.
Users may feel the urge to publish online the performances associated to
their last run -- including HR, as more and more products that monitor
it for fitness' sake are sold -- but experimenters and developers have
to remain cautious.

Physiological sensors are becoming cheaper and smaller, and hardware
manufacturers are increasingly interested in embedding them in their
products. With sensors acceptance, smartwatches may tomorrow provide a
wide range of continuous physiological data, along with remote sensing
through cameras. If users' rights and privacy are protected, this could
provide a wide range of areas for investigating and putting into
practice the similarity-attraction effect. Heart rate, galvanic skin
response, breathing, eye blinks: we ``classify'' events coming from the
outside world and it influences our physiology. An agent that seamlessly
reacts like us, based on the outputs we produce ourselves, could drive
users' engagement.

\begin{quote}
TODO: check refs, especially conference/journal titles + fairclough talk
+ original ref for uncanny valley?
\end{quote}

\balance

\bibliographystyle{apalike}
{
\small
\bibliography{biblio.bib}}

\begin{thebibliography}{}

\bibitem[Agelink et~al., 2001]{Agelink2001}
Agelink, M.~W., Malessa, R., Baumann, B., Majewski, T., Akila, F., Zeit, T.,
  and Ziegler, D. (2001).
\newblock {Standardized tests of heart rate variability: normal ranges obtained
  from 309 healthy humans, and effects of age, gender, and heart rate}.
\newblock {\em Clinical Autonomic Research}, 11(2):99--108.

\bibitem[Becker and Prendinger, 2005]{Becker2005}
Becker, C. and Prendinger, H. (2005).
\newblock {Evaluating affective feedback of the 3D agent max in a competitive
  cards game}.
\newblock In {\em Affective Computing and Intelligent Interaction}, pages
  466--473.

\bibitem[Berta et~al., 2013]{Berta2013}
Berta, R., Bellotti, F., {De Gloria}, A., Pranantha, D., and Schatten, C.
  (2013).
\newblock {Electroencephalogram and Physiological Signal Analysis for Assessing
  Flow in Games}.
\newblock {\em IEEE Transactions on Computational Intelligence and AI in
  Games}, 5(2):164--175.

\bibitem[Bestgen et~al., 2004]{Bestgen2004}
Bestgen, Y., Fairon, C., and Kerves, L. (2004).
\newblock {Un barometre affectif effectif: Corpus de r\'{e}f\'{e}rence et
  m\'{e}thode pour d\'{e}terminer la valence affective de phrases}.
\newblock {\em Journ\'{e}es internationales d’analyse statistique des
  donn\'{e}s textuelles (JADT)}.

\bibitem[Fairclough, 2014]{Fairclough2014}
Fairclough, S.~H. (2014).
\newblock {Human Sensors - Perspectives on the Digital Self}.
\newblock Keynote at Sensornet '14.

\bibitem[Fukushima et~al., 2011]{Fukushima2011}
Fukushima, H., Terasawa, Y., and Umeda, S. (2011).
\newblock {Association between interoception and empathy: evidence from
  heartbeat-evoked brain potential.}
\newblock {\em International journal of psychophysiology : official journal of
  the International Organization of Psychophysiology}, 79(2):259--65.

\bibitem[Gerlinghaus et~al., 2012]{Gerlinghaus2012}
Gerlinghaus, F., Pierce, B., Metzler, T., Jowers, I., Shea, K., and Cheng, G.
  (2012).
\newblock {Design and emotional expressiveness of Gertie (An open hardware
  robotic desk lamp)}.
\newblock {\em IEEE RO-MAN '12}, pages 1129--1134.

\bibitem[Harrison et~al., 2012]{Harrison2012}
Harrison, C., Horstman, J., Hsieh, G., and Hudson, S. (2012).
\newblock {Unlocking the expressivity of point lights}.
\newblock In {\em CHI '12}, page 1683, New York, New York, USA. ACM Press.

\bibitem[Huppi et~al., 2003]{Huppi2003}
Huppi, B.~Q., Stringer, C.~J., Bell, J., and Capener, C.~J. (2003).
\newblock {United States Patent 6658577: Breathing status LED indicator}.

\bibitem[ITU, 2003]{ITU2003}
ITU (2003).
\newblock {P. 851, Subjective Quality Evaluation of Telephone Services Based on
  Spoken Dialogue Systems}.
\newblock {\em International Telecommunication Union, Geneva}.

\bibitem[Karlesky and Isbister, 2014]{Karlesky2014}
Karlesky, M. and Isbister, K. (2014).
\newblock {Designing for the Physical Margins of Digital Workspaces: Fidget
  Widgets in Support of Productivity and Creativity}.
\newblock In {\em TEI '14}.

\bibitem[Kranjec et~al., 2014]{Kranjec2014}
Kranjec, J., Begu\v{s}, S., Ger\v{s}ak, G., and Drnov\v{s}ek, J. (2014).
\newblock {Non-contact heart rate and heart rate variability measurements: A
  review}.
\newblock {\em Biomedical Signal Processing and Control}, 13:102--112.

\bibitem[{Le Tallec} et~al., 2011]{LeTallec2011}
{Le Tallec}, M., Antoine, J.-Y., Villaneau, J., and Duhaut, D. (2011).
\newblock {Affective interaction with a companion robot for hospitalized
  children: a linguistically based model for emotion detection}.
\newblock In {\em 5th Language and Technology Conference (LTC'2011)}.

\bibitem[Lee and Nass, 2003]{Lee2003}
Lee, K.~M. and Nass, C. (2003).
\newblock {Designing social presence of social actors in human computer
  interaction}.
\newblock In {\em Proceedings of the conference on Human factors in computing
  systems - CHI '03}, number~5, page 289, New York, New York, USA. ACM Press.

\bibitem[Lee et~al., 2014]{Lee2014}
Lee, M., Kim, K., Rho, H., and Kim, S.~J. (2014).
\newblock {Empa talk}.
\newblock In {\em CHI EA '14}, pages 1897--1902, New York, New York, USA. ACM
  Press.

\bibitem[Lisetti and Nasoz, 2004]{Lisetti2004}
Lisetti, C. L.~t. and Nasoz, F. (2004).
\newblock {Using Noninvasive Wearable Computers to Recognize Human Emotions
  from Physiological Signals}.
\newblock {\em EURASIP J ADV SIG PR}, 2004(11):1672--1687.

\bibitem[MacDorman, 2005]{MacDorman2005}
MacDorman, K. (2005).
\newblock {Androids as an experimental apparatus: Why is there an uncanny
  valley and can we exploit it}.
\newblock {\em CogSci-2005 workshop: toward social mechanisms of android
  science}, 3.

\bibitem[Mandryk et~al., 2006]{Mandryk2006}
Mandryk, R., Inkpen, K., and Calvert, T. (2006).
\newblock {Using psychophysiological techniques to measure user experience with
  entertainment technologies}.
\newblock {\em Behaviour \& Information Technology}.

\bibitem[Matthews et~al., 2002]{Matthews2002}
Matthews, G., Campbell, S.~E., Falconer, S., Joyner, L.~a., Huggins, J.,
  Gilliland, K., Grier, R., and Warm, J.~S. (2002).
\newblock {Fundamental dimensions of subjective state in performance settings:
  Task engagement, distress, and worry.}
\newblock {\em Emotion}, 2(4):315--340.

\bibitem[M\"{o}ller et~al., 2007]{Moller2007}
M\"{o}ller, S., Smeele, P., Boland, H., and Krebber, J. (2007).
\newblock {Evaluating spoken dialogue systems according to de-facto standards:
  A case study}.
\newblock {\em Computer Speech \& Language}, 21(1):26--53.

\bibitem[Picard, 1995]{Picard1995}
Picard, R.~W. (1995).
\newblock {Affective computing}.
\newblock Technical Report 321, MIT Media Laboratory.

\bibitem[Prendinger et~al., 2004]{Prendinger2004}
Prendinger, H., Dohi, H., and Wang, H. (2004).
\newblock {Empathic embodied interfaces: Addressing users' affective state}.
\newblock In {\em Affective Dialogue Systems}, pages 53--64.

\bibitem[Reidsma et~al., 2010]{Reidsma2010a}
Reidsma, D., Nijholt, A., Tschacher, W., and Ramseyer, F. (2010).
\newblock {Measuring Multimodal Synchrony for Human-Computer Interaction}.
\newblock In {\em 2010 International Conference on Cyberworlds}, pages 67--71.
  IEEE.

\bibitem[Renard et~al., 2010]{Renard2010}
Renard, Y., Lotte, F., Gibert, G., Congedo, M., Maby, E., Delannoy, V.,
  Bertrand, O., and L\'{e}cuyer, A. (2010).
\newblock {OpenViBE: An Open-Source Software Platform to Design, Test, and Use
  Brain–Computer Interfaces in Real and Virtual Environments}.
\newblock {\em Presence: Teleoperators and Virtual Environments}, 19(1):35--53.

\bibitem[Slov\'{a}k et~al., 2012]{Slovak2012}
Slov\'{a}k, P., Janssen, J., and Fitzpatrick, G. (2012).
\newblock {Understanding heart rate sharing: towards unpacking physiosocial
  space}.
\newblock {\em CHI '12}, pages 859--868.

\bibitem[Walmink et~al., 2014]{Walmink2014}
Walmink, W., Wilde, D., and Mueller, F.~F. (2014).
\newblock {Displaying Heart Rate Data on a Bicycle Helmet to Support Social
  Exertion Experiences}.
\newblock In {\em TEI '14}.

\bibitem[Winton et~al., 1984]{Winton1984}
Winton, W.~M., Putnam, L.~E., and Krauss, R.~M. (1984).
\newblock {Facial and autonomic manifestations of the dimensional structure of
  emotion}.
\newblock {\em Journal of Experimental Social Psychology}, 20(3):195--216.

\bibitem[Wright and McCarthy, 2008]{Wright2008}
Wright, P. and McCarthy, J. (2008).
\newblock {Empathy and experience in HCI}.
\newblock {\em CHI '08}.

\end{thebibliography}

\end{document}